\documentclass[prl,twocolumn,amsfonts,aps, nofootinbib, showkeys,epsfigs]{revtex4}

   \usepackage{graphicx}
   \usepackage{graphics}

\usepackage{amsmath}
\usepackage{graphicx}
\usepackage{dcolumn}
\usepackage{bm}

\newcommand{\glog}{\Lambda}
\newcommand{\gexp}{{\cal E}}

\begin{document}

\title{Generalized (c,d)-entropy and aging random walks}

\author{Rudolf Hanel$^1$ and Stefan Thurner$^{1,2,3*}$}

\affiliation{$^1$Section for Science of Complex Systems, Medical University of Vienna, Spitalgasse 23, A-1090, Austria\\ 
$^2$Santa Fe Institute, 1399 Hyde Park Road, Santa Fe, NM 87501, USA\\
$^3$International Institute for Applied Systems Analysis, Schlossplatz 1, A-2361 Laxenburg, Austria}
\email{stefan.thurner@meduniwien.ac.at}

\begin{abstract}
Complex systems are often inherently non-ergodic and non-Markovian for which Shannon entropy loses its applicability. 
In particular accelerating, path-dependent, and aging random walks offer an intuitive picture for these non-ergodic and 
non-Markovian systems.  It was shown that the entropy of non-ergodic systems can still be derived from three of the 
Shannon-Khinchin axioms, and by violating the fourth -- the so-called composition axiom. The corresponding entropy is 
of the form $S_{c,d} \sim \sum_i \Gamma(1+d,1-c\ln p_i)$ and depends on two system-specific scaling exponents, $c$ and $d$. 
This entropy contains many recently proposed entropy functionals  as special cases, including Shannon and Tsallis entropy. 
It was shown that this entropy is relevant for a special class of non-Markovian random walks. In this work we generalize these walks
to a much wider class of stochastic systems that can be characterized as `aging' systems. 
These are systems whose transition rates between states are path- and time-dependent. We show that for particular aging walks 
$S_{c,d}$ is again the correct extensive entropy. 
Before the central part of the paper we review the concept of $(c,d)$-entropy in a self-contained way.


\end{abstract}

\maketitle

\section{Introduction - mini-review of $(c,d)$-entropy}

In their seminal works, Shannon and Khinchin showed that assuming four information theoretic axioms the entropy must be of Boltzmann-Gibbs  type, 
$S=-\sum_i p_i \log p_i$. In many physical systems one of these axioms may be violated. For non-ergodic systems the so called
separation axiom  (Shannon-Khinchin axiom 4) is not  valid. We show that whenever this axiom is 
violated the entropy takes a more general form, $S_{c,d}\propto \sum_i ^W \Gamma(d+1, 1- c \log p_i)$, where $c$ and $d$ are scaling exponents
and $\Gamma(a,b)$ is the incomplete gamma function. 
These exponents $(c,d)$ define equivalence classes for  {\em all}!,  interacting and non interacting, systems
and  unambiguously characterize any statistical system in its thermodynamic limit. 
The proof is possible because of two newly discovered scaling laws which  
any entropic form has to fulfill, if the first three Shannon-Khinchin axioms hold \cite{Hanel2011}. 
$(c,d)$ can be used to define equivalence classes of statistical systems. A series of known entropies  can be  classified in terms of these equivalence classes.
We show that the corresponding distribution functions are special forms of Lambert-${\cal W}$ exponentials containing  -- as  special cases  -- 
Boltzmann,  stretched exponential, and Tsallis distributions (power-laws).  
We go on by showing how the dependence of phase space volume $W(N)$ of a classical system on its size $N$, 
uniquely determines its extensive entropy, and in particular that the requirement of extensively fixes the exponents $(c,d)$,  \cite{Hanel2011_2}. 
We give a concise criterion when this entropy is not of 
Boltzmann-Gibbs type but has to assume a {\em generalized} (non-additive) form.  We showed  that generalized 
entropies can only exist when the dynamically (statistically) relevant fraction of degrees of freedom in the system 
vanishes in the thermodynamic limit \cite{Hanel2011_2}. These are systems where the bulk of the degrees of freedom is frozen 
and is practically statistically inactive. Systems governed by generalized entropies are therefore systems whose 
phase space volume effectively collapses to a lower-dimensional 'surface'. We explicitly illustrated the 
situation for binomial processes and argue that generalized entropies could be relevant for self organized 
critical systems such as sand piles,  for spin systems which form meta-structures such as vortices, domains, 
instantons, etc., and for problems associated with anomalous diffusion \cite{Hanel2011_2}. 
In this contribution we largely follow the lines of thought presented in \cite{Hanel2011,Hanel2011_2,ebook}. 

Theorem number 2 in the seminal 1948 paper, {\em The Mathematical Theory of Communication} \cite{shannon},
by Claude Shannon, proves the existence of the one and only form of entropy, given that 
three fundamental requirements hold. A few years later A.I. Khinchin remarked  in his {\em Mathematical Foundations of Information Theory} \cite{kinchin_1}: 
``However, Shannon's treatment is not always sufficiently complete and mathematically correct so that, besides having to free the theory from practical  
details, in many instances I have amplified and changed both the statement of definitions and the statement of proofs of theorems.''
Khinchin adds a fourth axiom. The three fundamental requirements of Shannon, in the `amplified' version of Khinchin,  are known as the Shannon-Khinchin (SK) axioms. 
These axioms list the requirements needed for an entropy to be a reasonable measure of the `uncertainty'  about a finite probabilistic system. 
Khinchin further suggests to also use entropy as a measure of the information {\em gained} about a system when making an 'experiment', i.e. by observing a realization 
of the probabilistic system. 

$\bullet$ Khinchin's first axiom states that for a system with $W$ potential outcomes (states) each of which is given by a probability $p_i\geq0$, with $\sum_{i=1}^W p_i=1$, 
the entropy $S(p_1, \cdots, p_W)$ as a measure of uncertainty about the system must take its maximum for the equi-distribution $p_i=1/W$, for all $i$.

$\bullet$ Khinchin's second axiom (missing in \cite{shannon}) states that any  entropy should remain invariant under adding zero-probability states to the system, 
i.e. $S(p_1, \cdots, p_W)=S(p_1, \cdots, p_W,0)$. 

$\bullet$ Khinchin's third axiom (separability axiom) finally makes a statement of the composition of two  finite probabilistic systems  $A$ and $B$. 
If the systems are independent of each other, entropy should be additive, meaning that the entropy of the combined system 
$A+B$ should be the sum of the individual systems, $S({A+B}) = S(A) + S(B)$. 
 If the two systems are dependent on each other, the entropy of the combined system,   
i.e. the information given by the realization of the two finite schemes $A$ and $B$, $S(A+B)$,  is equal to the 
information gained by a realization of system $A$,  $S(A)$, plus the mathematical expectation of information gained
 by a realization of system $B$, after the realization of system $A$, $S({A+B}) = S(A) + S|_A(B)$. 

$\bullet$ Khinchin's fourth axiom is the requirement  that entropy is a continuous function of all its arguments $p_i$ and does not depend 
on anything else.

Given these axioms, the {\em Uniqueness theorem} \cite{kinchin_1} states that the one and only possible entropy is 
\begin{equation}
S(p_1,\cdots , p_W) = -k \sum_{i=1}^{W}p_i\log p_i \quad , 
\end{equation}
where $k$ is an arbitrary positive constant. The result is of course the same as Shannon's. 
We call the combination of 4 axioms  the Shannon-Khinchin (SK) axioms. 

From information theory now to physics, where systems may exist that violate the separability axiom. This might 
especially be the case for non-ergodic, complex systems exhibiting long-range  and strong interactions. 
Such complex systems may show extremely rich behavior in contrast to simple ones, such as gases.  
There exists some hope that it should be possible to understand such systems also on a thermodynamical basis, meaning 
that  a few measurable quantities would be sufficient to understand their macroscopic phenomena. 
If this  would be possible, through an equivalent to the  second law of thermodynamics, some appropriate entropy would enter  
as a fundamental concept relating the number of microstates in the system to  its macroscopic properties.  
Guided by this hope, a series of so called generalized entropies have been suggested over the past decades, 
see \cite{tsallis88, celia, kaniadakis,curado,expo_ent,ggent} and Table 1. 
These entropies have been designed for different purposes and have not been related to a fundamental origin. 
Here we ask how generalized entropies can look like if they fulfill some of  the Shannon-Khinchin axioms, 
but explicitly violate the separability axiom. We do this  axiomatically as first presented in \cite{Hanel2011}. 
By doing so we can relate a large class of generalized entropies to a single fundamental origin. 

The reason why  this axiom is violated in some physical, biological or social systems is {\em broken ergodicity}, 
i.e. that not all regions in phase space are visited and many micro states are effectively `forbidden'. 
Entropy relates the number of micro states of a system to an {\em extensive} quantity, which plays the 
fundamental role in the systems thermodynamical description. Extensive means that if two initially isolated, i.e. sufficiently separated 
systems, $A$ and $B$, with $W_A$ and $W_B$ the respective numbers of states, are brought together, 
the entropy of the combined system 
$A+B$ is $S(W_{A+B}) = S(W_A) + S(W_B)$. 
$W_{A+B}$ is the number of states in the combined system $A+B$.
This is not to be confused with {\em additivity} which is the property that $S(W_A W_B) = S(W_A) + S(W_B)$. 
Both, extensivity and additivity coincide if  number of states in the combined system is $W_{A+B}=W_AW_B$.
Clearly,  for a non-interacting system Boltzmann-Gibbs-Shannon entropy, $S_{\rm BG}[p]= - \sum_i^W   p_i\ln p_i$,  is extensive {\em and} additive.
By 'non-interacting'  (short-range, ergodic, sufficiently mixing, Markovian, ...) systems we mean  $W_{A+B}=W_AW_B$. 
For interacting statistical systems the latter is in general not true; phase space is only partly visited and $W_{A+B} < W_AW_B$. In this case,   
an additive entropy  such as Boltzmann-Gibbs-Shannon can no longer be  extensive and vice versa. 
To ensure extensivity of entropy, an entropic form should be found for the particular interacting statistical systems at hand. 
These entropic forms are called {\em generalized entropies}  and usually assume trace form \cite{tsallis88, celia, kaniadakis,curado,expo_ent,ggent}
\begin{equation}
 S_g[p]=\sum_{i=1}^W g(p_i) \quad ,
\label{S_g} 
\end{equation} 
$W$ being the number of states. 
Obviously not all generalized entropic forms are of this type. R\'enyi entropy e.g. is  
of the form $G(\sum_{i}^W g(p_i))$, with $G$ a monotonic function.
We use trace forms Eq. (\ref{S_g}) for simplicity. R\'enyi forms can be studied in exactly the same way 
as will be shown, however at more technical cost. 

Let us revisit the Shannon-Khinchin axioms in the light of generalized entropies of trace form   Eq. (\ref{S_g}). 
Specifically axioms SK1-SK3 (now re-ordered) have implications on  the functional form of $g$

\begin{itemize}
\item
SK1: The requirement that $S$ depends continuously on $p$  implies that $g$ is a continuous function.
\item
SK2: The requirement that the entropy is maximal for the equi-distribution $p_i=1/W$ 
(for all $i$)  implies that $g$ is a concave function.
\item 
SK3: The requirement that adding a zero-probability state to a system, $W+1$ with $p_{W+1}=0$,  
does not change the entropy, implies that $g(0)=0$. 
\item 
SK4 (separability axiom): The entropy of a system -- composed of sub-systems $A$ and $B$ -- equals the entropy of $A$ plus the expectation value of the entropy 
of $B$, conditional on $A$. Note that this also corresponds exactly to Markovian processes.
\end{itemize}
As mentioned, if SK1 to SK4 hold, the only possible entropy is the Boltzmann-Gibbs-Shannon entropy. 
We are now going to derive the extensive entropy when the separability axiom SK4 is violated. 
Obviously this entropy will be more general and should contain BG entropy as a special case. 

We now assume that axioms SK1, SK2, SK3  hold, i.e. we restrict ourselves to trace form entropies with $g$ continuous, concave 
and $g(0)=0$. These systems we call {\em admissible} systems. Admissible systems when combined  with a maximum 
entropy principle show remarkably simple mathematical properties  \cite{Hanel2011_b,Hanel2012_b}.

This generalized entropy for  (large) admissible statistical systems (SK1-SK3 hold)   
is derived from two  hitherto unexplored fundamental scaling laws of extensive entropies \cite{Hanel2011}. 
Both scaling laws are characterized by  exponents $c$ and $d$, respectively, which allow to uniquely define equivalence classes  of entropies,  
meaning that two entropies are equivalent in the thermodynamic limit 
if their exponents $(c,d)$ coincide. Each admissible system belongs to  one of these equivalence classes $(c,d)$, \cite{Hanel2011}. 

In terms of the exponents $(c,d)$ we showed in \cite{Hanel2011} that all generalized entropies  have the form 
\begin{equation}
S_{c,d}\propto\sum_i ^W  \Gamma(d+1, 1- c \log p_i)
\label{gent}
\end{equation}
with  $\Gamma(a,b)=\int_b^\infty dt\,t^{a-1}\exp(-t)$ the incomplete Gamma-function.

\begin{figure}[t]
 \begin{center}
 	\includegraphics[width=\columnwidth]{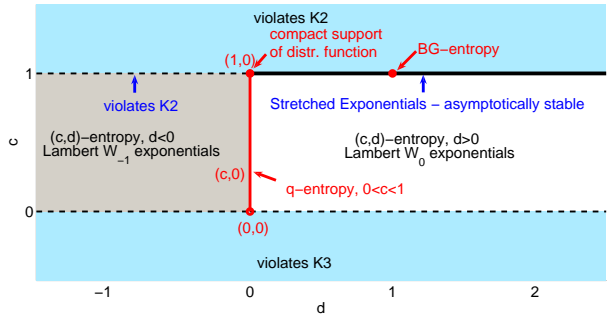}
 \end{center}
\caption{
Entropies parametrized in the $(c,d)$-plane, with their associated distribution functions. 
BG entropy corresponds to $(1,1)$, Tsallis entropy to $(c,0)$, and entropies for stretched exponentials to $(1,d>0)$.
Entropies leading to distribution functions with compact support, 
belong to equivalence class $(1,0)$. Figure from \cite{ebook}. 
\label{classfig}
}
\end{figure}

\subsubsection{Special cases of equivalence classes}
Let us look at some specific equivalence classes $(c,d)$
\begin{itemize}
\item Boltzmann-Gibbs entropy belongs to the $(c,d)=(1,1)$ class. One gets from Eq. (\ref{gent})  
 	\begin{equation}
  	S_{1,1}[p]= \sum_i g_{1,1}(p_i)=   -\sum_i p_i\ln p_i +1  
  	\quad . 
 	\end{equation}
	
\item Tsallis entropy belongs to the  $(c,d)=(c,0)$ class. From Eq. (\ref{gent}) and the choice $r=1/(1-c)$ (see below)
	we get
 	\begin{equation}
		\begin{array}{lcl}
	 		S_{c,0}[p] = \sum_i g_{c,0}(p_i)=  \frac{1-\sum_i p_i^c}{c-1} +1 \, . 
		\end{array}
 	\end{equation}
	Note, that although the {\em pointwise} limit $c\to 1$ of Tsallis entropy yields  BG entropy, the asymptotic properties $(c,0)$ do {\em not} 
	change continuously to $(1,1)$ in this limit! In other words the thermodynamic limit and the limit $c\to 1$ do not commute.

\item The entropy related to  stretched exponentials \cite{celia} belongs  to the $(c,d)=(1,d)$ classes, see Table 1.
	As a specific example  we compute the $(c,d)=(1,2)$ case,  
	\begin{equation}
  		S_{1,2}[p]= 2  \left(1-\sum_i p_i \ln p_i  \right) + \frac{1}{2}\sum_i p_i\left(\ln p_i  \right)^2, 
	\label{c1d2}
 	\end{equation}
 	leading to a superposition of two entropy terms, the asymptotic behavior being dominated by the second.  	
\end{itemize}
Other entropies which are special cases of our scheme are found in Table 1. 

Inversely, for any given entropy we are now in the remarkable position to characterize {\em all} 
large SK1-SK3 systems by a pair of two exponents $(c,d)$, see Fig. \ref{classfig}.
For example, for $g_{\rm BG}(x)=-x\ln(x)$ we have  $c=1$, 
and  $d=1$. $S_{\rm BG}$ therefore belongs to the universality class $(c,d)=(1,1)$.  
For $g_{q}(x)= (x-x^q)/(1-q)$ (Tsallis entropy) and $0<q<1$ one finds 
 $c=q$ and $d=0$, and Tsallis entropy, $S_{q}$, belongs to the universality class $(c,d)=(q,0)$. 
Other examples are listed in Table 1. 

The universality classes $(c,d)$ are equivalence classes with the 
equivalence relation given by:  
$g_{\alpha} \equiv g_{\beta} \Leftrightarrow c_\alpha=c_\beta$ and $d_\alpha=d_\beta$.
This relation partitions the space of all admissible $g$ into equivalence classes completely specified by the pair $(c,d)$.  

\begin{table*}[t]
\caption{
Order in the zoo of recently introduced entropies for which  SK1-SK3 hold. All of them  are  special cases of the entropy given in Eq. (\ref{gent}) and their  asymptotic behavior 
is uniquely determined by $c$ and $d$. 
It can be seen immediately that $S_{q>1}$, $S_{b}$ and $S_{E}$ are asymptotically identical; so are $S_{q<1}$ and $S_{\kappa}$, as well as $S_{\eta}$ and $S_{\gamma}$.  
}
\centering
{\small
\begin{tabular}{ll|c|c|l}
entropy &  & $c$ & $d$ & reference \\ \hline
$S_{c,d} = er \sum_i \Gamma(d+1, 1-c\ln p_i) - cr $ 	& $(r=(1-c+cd)^{-1})$								& $c$ 		& $d$  			&  \\  \hline
$S_{BG} =  \sum_i p_i  \ln  (1/p_i)    $ 			&												& $1$ 		& $1$  			&   \cite{kinchin_1} \\  \hline
$S_{q<1}(p) = \frac{1-\sum{p_i^q}}{q-1}$ 			& $(q<1)$ 										& $c=q<1$ 	& $0$  			&  \cite{tsallis88} \\  \hline
$S_{\kappa}(p) = - \sum_i p_i \frac{p_i^{\kappa}  - p_i^{-\kappa}  }{2\kappa}  $ 	 &	($0<\kappa \leq1$)						& $c=1-\kappa$ & $0$  			&   \cite{kaniadakis} \\  \hline
$S_{q>1}(p) = \frac{1-\sum{p_i^q}}{q-1}$ 			& $(q>1)$ 										& $1$ 		& $0$  			&   \cite{tsallis88} \\  \hline
$S_{b}(p) = \sum_i (1-e^{-bp_i})   + e^-b -1$ 		& $(b>0)$ 										& $1$ 		& $0$  			&   \cite{curado} \\  \hline
$S_{E}(p) =  \sum_i p_i (1-e^{\frac{p_i -1}{p_i}})    $  &												& $1$ 		& $0 $  			&   \cite{expo_ent} \\  \hline
$S_{\eta}(p) = \sum_i  \Gamma(\frac{\eta+1}{\eta},-\ln p_i) - p_i \Gamma (\frac{\eta+1}{\eta})$  &$ (\eta>0)$ 		& $1$ 		& $d=\frac{1}{\eta}$  	&   \cite{celia} \\  \hline
$S_{\gamma}(p) =  \sum_i p_i \ln ^{1  / \gamma} (1/p_i)    $ 	&										& $1$ 		& $d =1/\gamma $  	&   \cite{TsallisBook_2009} 
\\  \hline
$S_{\beta}(p) =  \sum_i p_i^{\beta} \ln  (1/p_i)    $ 	&												& $c=\beta$  	& $1$  	&    \cite{shafee} \\  \hline
\end{tabular}
}
\end{table*}

\subsection{Distribution functions}

Distribution functions associated with  our  $\Gamma$-entropy, Eq. (\ref{gent}), 
can be derived from so-called generalized logarithms of the entropy. Under the 
maximum entropy principle (given ordinary constraints) the inverse functions of these logarithms, $\gexp=\glog^{-1}$, 
are the distribution functions, 
$p(\epsilon) = \gexp_{c,d,r}(-\epsilon)$, where 
for example $r$ can be chosen $r=(1-c+cd)^{-1}$. 
One finds \cite{Hanel2011}
\begin{equation}
\gexp_{c,d,r}(x)=  e^{    - \frac{d}{1-c}  \left[ {\cal W}_k \left( B (1-x/r )^{ \frac{1}{d} } \right)  -  {\cal W}_k(B)  \right]  }\, ,
\label{gexp}
\end{equation}
with the constant $B\equiv \frac{(1-c)r}{1-(1-c)r} \exp \left(  \frac{(1-c)r}{1-(1-c)r} \right) $. 
The function ${\cal W}_k$ is the $k$'th branch of the Lambert-${\cal W}$ function which -- as a solution to the equation
$x={\cal W}(x)\exp({\cal W}(x))$ -- has only two real solutions $W_k$, the branch $k=0$ and branch $k=-1$. 
Branch $k=0$ covers  the classes for $d\geq 0$, branch $k=-1$ those for $d<0$.

\subsubsection{Special cases of distribution functions}

It is easy to verify that the class $(c,d)=(1,1)$ leads to Boltzmann distributions, and the class $(c,d)=(c,0)$ yields 
power-laws, or more precisely, Tsallis distributions i.e.  $q$-exponentials. 
All classes associated with  $(c,d)=(1,d)$, for $d>0$ are associated with stretched exponential distributions.   
Expanding the $k=0$ branch of the Lambert-${\cal W}$ function  $W_0(x)\sim x-x^2+\dots$ for $1\gg|x|$, 
the limit $c\to 1$ is shown to be a stretched exponential. 
It was shown that $r$ does not effect its asymptotic properties (tail of the distributions), but can be used to incorporate  
finite size properties of the distribution function for small $x$.  

\subsection{How to determine the exponents $c$ and $d$?}

In \cite{Hanel2011_2} we have shown that the requirement of extensivity determines uniquely both exponents $c$ and $d$.
What does extensivity mean?
Consider a system with $N$ elements. The number of system configurations 
(microstates) as a function of $N$ are denoted by $W(N)$. 
Starting with SK2, $p_i=1/W$ (for all $i$), we have  $S_g=\sum_{i=1}^{W} g(p_i) = W g(1/W)$.  
As mentioned above extensivity for two subsystems $A$ and $B$ means that
\begin{equation}
	W_{A+B}g\left( 1/W_{A+B}  \right) = W_{A}g\left( 1/ W_{A}  \right) + W_{B}g\left( 1/ W_{B}  \right) \quad .
\label{dum1}
\end{equation}
Using this equation one can straight forwardly derive the formulas  (for details see \cite{Hanel2011_2})
\begin{equation}
	\frac{1}{1-c}= \lim_{N \to \infty} N \frac{ W'(N)}{W(N)} \quad .
	 \label{cformel}
\end{equation}
\begin{equation}
	d=  \lim_{N \to \infty}  \log  W  \left( \frac1N \frac{W}{ W'} +c-1 \right) \quad .
\label{dformel}
\end{equation}
Here $W'$ means the derivative with respect to $N$.

\subsection{A note on R\'enyi-type entropies}

R\'enyi entropy is obtained by relaxing SK4 to the  unconditional additivity condition. 
Following the same scaling idea for R\'enyi-type entropies,  $S=G(\sum_{i=1}^W g(p_i))$, with $G$ and $g$ some functions, one gets 
\begin{equation}
\lim_{W\to\infty} \frac{S(\lambda W)}{S(W)}= \lim_{s\to\infty} \frac{ G \left( \lambda f_g(\lambda^{-1})s \right)  }{G(s)} \quad ,
\end{equation} 
where $f_g(z)=\lim_{x\to 0} g(zx)/g(x)$. 
The expression $f_G(s)\equiv \lim_{s} G(s y)/G(s)$, 
provides the starting point for deeper analysis which now gets more involved. 
In particular, for R\'enyi entropy with $G(x) \equiv \ln(x)/(1-\alpha)$ and $g(x)\equiv x^{\alpha}$, 
the asymptotic properties yield the class $(c,d)=(1,1)$, (BG entropy) meaning that R\'enyi entropy is additive. 
However, in contrast to the trace form entropies used above, 
R\'enyi entropy can be shown to be {\em not} Lesche stable, as was observed before \cite{lesche,Abe_2002, Arimitsu,Kaniadakis_2004,HT_robustness_1}. 
All of the $S=\sum_i^W g(p_i)$ entropies can be shown to be Lesche stable, see \cite{ebook}.

\section{Aging random walks}

In \cite{Hanel2011_2} we have discussed a particular type of an accelerating random walk that requires generalized entropy. 
We first revisit the example of this {\em auto-correlated} random walk $x$ and point out that all moments of this random walk 
are identical to the moments of an {\em accelerating} random walk. This means that two processes, where the the first  
requires a generalized entropy and the second requires  Shannon entropy, they both have the same distribution function asymptotically.
We then show that  auto-correlated random walks are asymptotically equivalent to {\em aging} random walks.

Random walks of length $N$ consist of sequences of $N$ decisions $\omega_n$ with $n=1,2,\cdots,N$. Each decision 
determines whether to take a step of size $\Delta x$ to the left, $\omega_n = -1$, or to the right, $\omega_n =1$, at time 
$t=n\Delta t$, with a  probability $q_+$ and $q_-$. The path $x(N\Delta t)$ is given by
\begin{equation}
x(N\Delta t)=\sum_{n=1}^N\omega_n\Delta x \quad .
\end{equation} 
In the following we  set the time increment $\Delta t=1$ and the step size $\Delta x=1$.

For the usual random walk each decision $\omega_n$ has no bias for any direction, i.e. $q_{+} =q_{-} =1/2$ and the expectation value 
$\langle\omega_n\rangle=q_{+} -q_{-} =0$. 
Further, all decisions are independent, meaning that $\langle \omega_m \omega_n \rangle=\delta_{nm}$, where $\delta_{nm}$ is the Kronecker delta. 
The number of possible paths $W$ such a random walk can take -- its phase-space volume for $N$ steps -- is given by $W(N)=2^N$. 
Using Eq. (\ref{cformel}) and Eq. (\ref{dformel}) one immediately finds $(c,d)=(1,1)$. 
Random walks consisting of independent decisions are described by Shannon's entropy.

\subsection{Accelerating and auto-correlated random walks}

In \cite{Hanel2011_2} we considered a different type of random walk where again decisions have no a 
priori bias on the direction of the walk, i.e. $\langle\omega_n\rangle=0$. However, decisions $\omega_n$ 
and $\omega_m$ are not  independent anymore. 
In particular we considered a constant $0<\alpha\leq 1$ such that
\begin{equation}
\langle\omega_m\omega_n\rangle=1\quad {\rm if}\quad z\leq n^\alpha,\,m^\alpha<z+1
\label{autocorcond}
\end{equation} 
for some $z=0,1,2,\cdots$, and $\langle\omega_m\omega_n\rangle=0$ otherwise. 
This means that the process is correlated with its history, and that after $n$ steps 
the number of free decisions is given by $z \sim n^{\alpha}$. As the walk progresses it 
heads persistently in the same direction for approximately $\frac1\alpha n^{1-\alpha}$ steps at the $n$'th step. 

\begin{figure}[t]
 \begin{center}
 	\includegraphics[width=0.75\columnwidth]{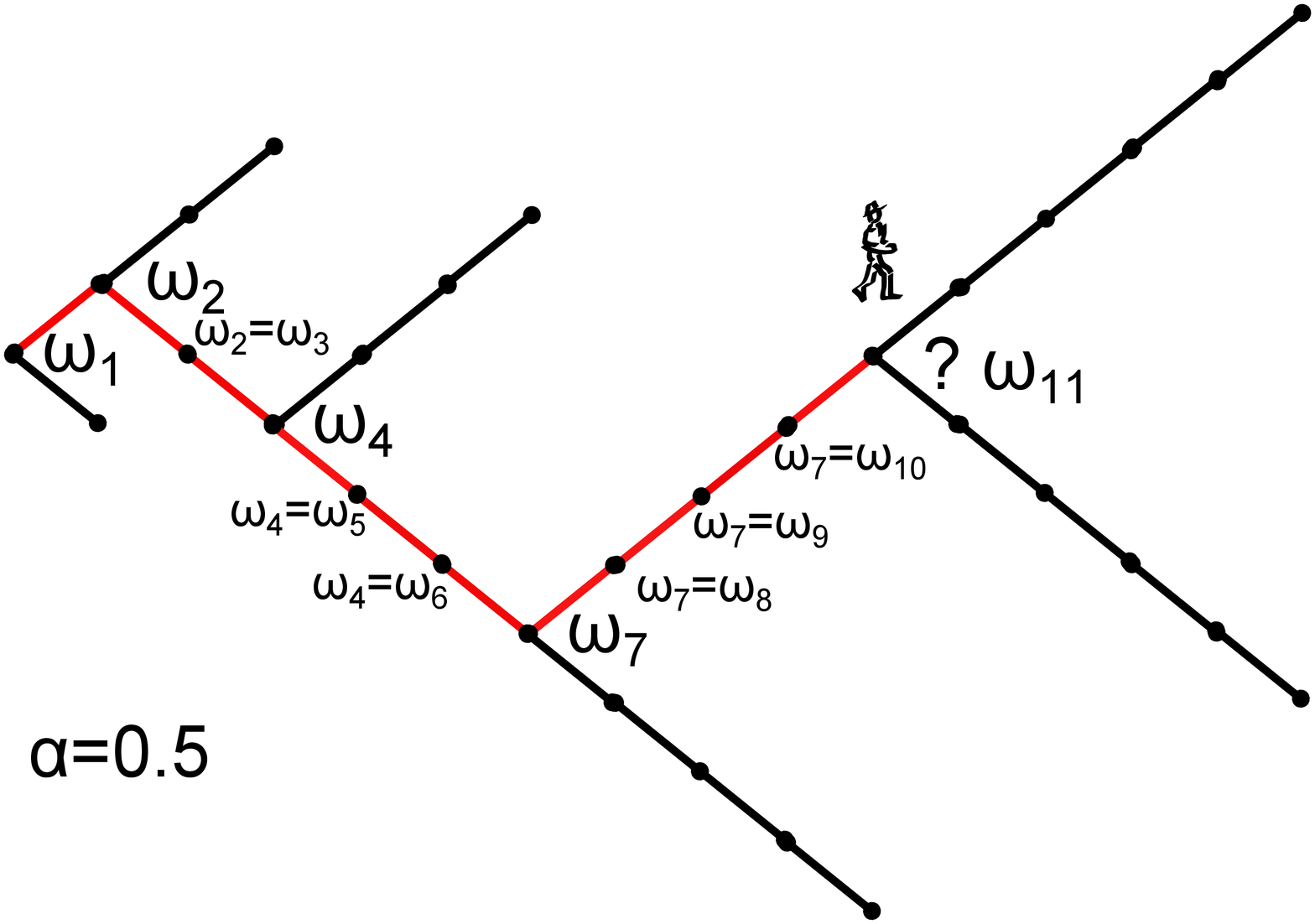}
 \end{center}
\caption{
Example for an auto-correlated random walk that persistently walks in the same direction for $\propto n^{1-\alpha}$ steps ($\alpha=0.5$). 
\label{fig:walk}
}
\end{figure}

Therefore, the number of possible paths $W$ grows like $W(N)=2^{N^{\alpha}}$ and the random walk has a 
stretched exponential growth of phase-space volume. Using Eq. (\ref{cformel}) and Eq. (\ref{dformel}), 
the universality class of the process belongs to $(c,d)=(1,\frac{1}{\alpha})$. 
Increasing persistence of a process over time therefore can be seen as the hallmark of 
processes that follow generalized extensive entropies.

Computing the moments of $x$ one finds that odd moments vanish, $\langle x^{2r+1}(N)\rangle=0$, where $r$ is a natural number, 
and even moments behave as
\begin{equation}
\langle x^{2r}(N)\rangle = \frac{(2r)!}{r!2^r} N^{(2-\alpha)r} \quad .
\end{equation}
The {\em auto-correlated} random walk therefore possesses the same moments as an {\em accelerated} random walk, 
i.e. a random walk with independent decisions 
$\langle\omega_m\omega_n\rangle=\delta_{nm}$, however with a time-dependent step size $\Delta x(n)=D(n)\Delta x$, 
that increases proportional to $n^{(1-\alpha)/2}$. Here  $D(n)$ is the time-dependent {\em `diffusion constant'} of the process. 
In particular, the second moment is given by
\begin{equation}
\langle x^2(N)\rangle = N^{2-\alpha} \quad .
\end{equation}
We  conclude that observable distribution functions do not necessary tell us which entropy class the process belongs to.
In this example the auto-correlated random walk of class $(c,d)=(1,1/\alpha)$ has all moments in common with the 
accelerated random walk, which is of class $(c,d)=(1,1)$.

\subsection{Generalization to aging (path-dependent) random walks}

The above generating rule Eq. (\ref{autocorcond}), for incorporating auto-correlations into random walks is somewhat artificial. 
We now show that it is possible to get a completely analogous auto-correlated behavior by considering {\em aging} in the 
decision process $\omega$. This can be done as follows. 
Consider a second process $\eta_n$, such that $\omega_{n}=\eta_n\omega_{n-1}$. 
This process indicates whether at step $n$ the random walk will proceed in the direction of the 
previous time step ($\eta_n=1$) or whether the walk reverses direction ($\eta_n=-1$). 
Let $k_+(N)$ ($k_-(N)$) be the number of times that $\eta_n=+1$ ($\eta_n=-1$) for 
$1\leq n\leq N$, i.e. $k=(k_+,k_-)$ is the histogram of the process $\eta$ up to time step $N$.  
Aging can now be incorporated by considering conditional probabilities for 
reversing direction or not. In particular we have
\begin{eqnarray}
p(\eta_{n+1}=1|k(n))=\frac{1}{1+\alpha [1+k_+(n)]^{\alpha-1}}\quad ,\nonumber \\
p(\eta_{n+1}=-1|k(n))=\frac{\alpha [1+k_+(n)]^{\alpha-1}}{1+\alpha [1+k_+(n)]^{\alpha-1}}\quad,
\label{transprob}
\end{eqnarray}
where $0<\alpha\leq 1$ takes the same numerical values as in the auto-correlated random walk. 
As a consequence these {\em aging} random walks are non-Markovian processes with memory, 
since the conditional probabilities for making the next decision depend on the entire history of the process. 
The dependence is such that the process conditions its next decisions on the histogram of decisions 
made in the past, not on its precise trajectory and again the decisions become increasingly 
persistent. To handle this type of process analytically is difficult. However, we can demonstrate 
numerically, that the first three even moments $\langle x^2\rangle$, $\langle x^4\rangle$, and 
$\langle x^6\rangle$ of the auto-correlated and the aging random walk are identical,
and also the number of reversal decisions $k_-$ of both processes asymptotically behave in exactly 
the same way. This shows that the effective number of different paths, i.e. the phase-space volume, of both 
processes grows in the same way and therefore the aging random walk belongs to the  
equivalence class $(c,d)=(1,1/\alpha)$. 

It is possible to show that
one can arrive at different equivalence class by altering the expression $n^{\alpha}$ in Eq. (\ref{autocorcond}). 
In particular by exchanging $n^{\alpha}$ with  $a \log n$ (same for $m$), one
arrives at the Tsallis equivalence class $(c,d)=(q,0)$.

\begin{figure}[t]
 \begin{center}
 	\includegraphics[width=1.1\columnwidth]{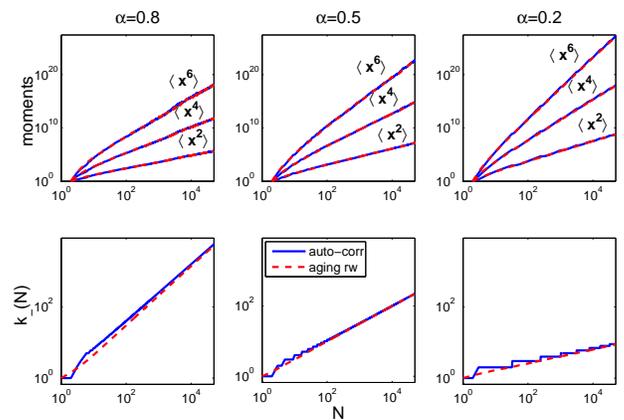}
 \end{center}
\caption{
Comparison of the first three even moments $\langle x^2(N)\rangle$, $\langle x^4(N)\rangle$, 
and $\langle x^6(N)\rangle$ and the average number of direction reversal-decisions $k_-(N)$, 
with $1\leq N\leq 50000$ for the auto-correlated random walk (blue lines) 
and aging random walks (red dashed lines) for values $\alpha=0.2$, $0.5$, and $0.8$.
\label{fig:comparison}
}
\end{figure}
  
\subsection{General classes of aging random walks}

\begin{figure}[t]
 \begin{center}
 	\includegraphics[width=\columnwidth]{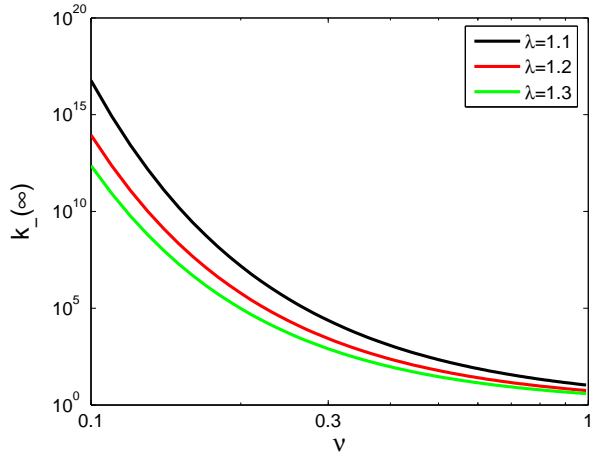}
 \end{center}
\caption{
The maximal number direction reversal decisions in random walks in entropic classes $(c,d)$ with $0<c<1$ for
the values $\lambda=1.1$, $1.2$, and $1.3$.
\label{fig:maxnum}
}
\end{figure}

We are now in the position to generalize random aging walks to different classes $(c,d)$ of entropies.
This can be done by generalizing the path dependent conditional probabilities of Eq. (\ref{transprob})
in the following way:
\begin{eqnarray}
p(\eta_{n+1}=1|k(n))=\frac{1}{1+g(k_+(n))}\quad, \nonumber  \\ 
p(\eta_{n+1}=-1|k(n))=\frac{g(k_+(n))}{1+g(k_+(n))}\quad,
\label{transprob2}
\end{eqnarray}
where $g(k_+)$ is a monotonically decreasing function ($\lim_{k_+\to\infty}g(k_+)=0$).
In the above example $g(k_+)=\alpha [1+k_+(n)]^{\alpha-1}$ corresponds to an aging process in the entropy class $(c,d)=(1,1/\alpha)$.
Different choices of the function $g$ will in general lead to different entropy classes $(c,d)$ depending on the asymptotic behavior 
$k_-(N)$ which corresponds to the effective number of free decisions occurring during the walk and therefore to the way phase-space grows with $N$. 
Again, a precise analytical analysis of how the choice of $g$ determines $(c,d)$ is complicated and goes beyond the scope of the paper.
However, it is known that systems with $0<c<1$ allow only a finite effective number of free decisions, e.g. \cite{Hanel2011_2,TMGMS}. 
This can for instance be achieved with the function
\begin{equation}
g(k_+)=\lambda^{-(k_+^\nu)}\,,
\end{equation}
with $0<\nu\leq 1$ and $\lambda>1$.
By using a `mean field' approach and setting 
\begin{equation}
\frac{dk_+(N)}{dN}=p(+1|k(N))\quad{\rm and}\quad\frac{dk_-(N)}{dN}=p(-1|k(N))
\end{equation}
one can derive the following asymptotic expression:
\begin{equation}
k_-=\frac{1}{\nu}(\log\lambda)^{-\frac{1}{\nu}}\gamma\left(\frac{1}{\nu},k_+^\nu\log\lambda\right)\,,
\end{equation}
where $\gamma(a,b)=\int_0^b dt t^{a-1}e^{-t}$ is the lower incomplete gamma function. Consequently the effective number 
of free decisions in this aging walks can be estimated by $k_-(\infty)$. The behavior of $k_-(\infty)$ is shown in Fig. (\ref{fig:maxnum}).

\begin{figure}[t]
 \begin{center}
 	\includegraphics[width=1.1\columnwidth]{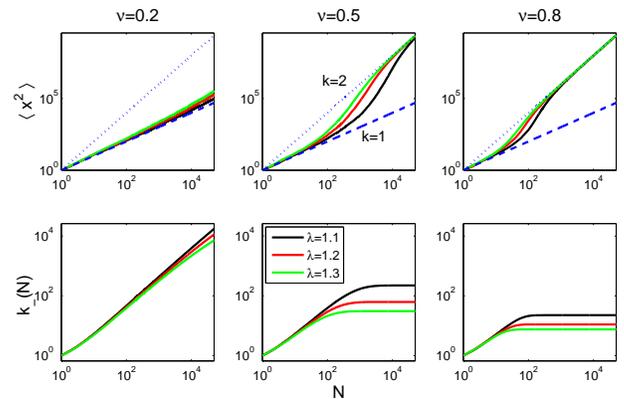}
 \end{center}
\caption{
In the three top panes the second moment $\langle x^2(N)\rangle$ is shown for $\nu=0.2$, $0.5$, and $0.8$, for  
$\lambda=1.1$ (black), $1.2$ (red), and $1.3$ (green).
The blue dotted and dashed lines indicate the function $N^2$ and $N$, respectively. 
A cross over from $\langle x^2(N)\rangle\sim N$ to $\langle x^2(N)\rangle\sim N^2$ is clearly visible for
$\nu=0.5$ and $0.8$. The three bottom panes show the average number of direction reversal-decisions $k_-(N)$. 
Simulations were performed in the range $1\leq N\leq 50000$. For $\nu \to1$ the crossover happens at smaller $N$, for all values of 
$\lambda$. 
\label{fig:crossover} 
}
\end{figure}

The fact that only a finite number, $k_-$, of direction reversal decisions happen during such a random walk 
leads to a peculiar cross-over phenomenon that can be observed by studying the second moment $\langle x^2(N)\rangle$ of the walk.
In particular $\langle x^2(N)\rangle\sim N$ for small $N$. For large $N\gg 1$ the random walk persistently heads into one direction and 
$\langle x^2(N)\rangle\sim N^2$. At an intermediate range of $N$ that depends on the value of $\lambda$ the behavior $\langle x^2(N)\rangle$
crosses over from $N$ to $N^2$, see Fig. (\ref{fig:crossover}). The derive
the exact function that relates $\nu$ and $\lambda$ to $c$ and $d$ is beyond the scope of this paper. 
However, we conjecture that $c=1-\nu$ since $\nu=0$ corresponds to the usual random walk and therefore we require $c=1$ in this case.

It would be desirable to have a comprehensive classfication of aging random walks in terms of 
equivalence classes $(c,d)$. We conjecture that this is in fact possible by exchanging the expression 
$n^{\alpha}$ in Eq. (\ref{autocorcond}) with more general forms $n^{\alpha} \to n^{\alpha} (\log n)^{\beta}$, 
where $\alpha$ and $\beta$ are directly related to $c$ and $d$.

Finally, let us remark that it is not straight forward to relate aging random walks and its class $(c,d)$ with  more traditional 
scaling exponents such as for example the Hurst exponent. The very nature of aging walks is that their persistence changes over time. 

\section{Conclusions}

Based on  recently discovered scaling laws for trace form entropies we can classify 
all statistical systems and assign the a unique system-specific (extensive) generalized entropy. 
For non-ergodic systems these entropies may deviate from the Shannon form. 
The exponents for BG systems are $(c,d)=(1,1)$, systems characterized by stretched exponentials  
belong to the class $(c,d)=(1,d)$, and Tsallis systems have $(c,d)=(q,0)$. 
A further interesting feature all admissible systems is that they  are all {\em Lesche stable}, and that 
the classification scheme for generalized entropies of type $S=\sum_i g(p_i)$ can be easily 
extended to entropies of R\'enyi type, i.e. $S=G(\sum_{i} g(p_i))$. For proofs see \cite{ebook}.

We  demonstrated that the auto-correlated random walk characterized by $0<\alpha\leq1$ introduced 
in \cite{Hanel2011_2} can not be distinguished from accelerating random walks. Although the presented auto-correlated 
random walk is of entropy class $(c,d)=(1,1/\alpha)$ and the accelerated random walk is 
of class $(c,d)=(1,1)$, both processes have the same distribution function since all moments $\langle x^n\rangle$ 
are identical.  
We have shown that other classes of random walks can  naturally be obtained, including those belonging to the 
$(c,d)=(q,0)$, or Tsallis equivalence class. 
Moreover, we showed numerically that the auto-correlated random walk is asymptotically equivalent to a 
particular {\em aging} random walks, where the probability of a decision to reverse the direction of the walk 
depends on the path the random walk has taken. This concept of aging can easily be generalized to different 
forms of aging and it can be expected that many of the admissible systems can be represented 
by a specific type of aging that is specified by the aging function $g$, Eq. (\ref{transprob2})).
Finally, we have seen that different equivalence classes $(c,d)$ can be realized by specifying a 
aging function $g$. The effective number of direction reversal decisions corresponding to the aging function remains finite 
and therefore the associated generalized entropy requires a class $(c,d)$ with $0<c<1$. 
We believe that it should be possible that the scheme of aging random walks can be naturally 
extended to aging processes in physical, biological, and social systems in general.







\bibliographystyle{mdpi}

\begin{thebibliography}{1}




\bibitem{Hanel2011} 
Hanel R.; Thurner S.
 A comprehensive classification of complex statistical systems and an axiomatic derivation of their entropy and distribution functions.
 {\em Europhys Lett}  {\bf 2011} {\em 93}, 20006.

\bibitem{Hanel2011_2} 
Hanel R.; Thurner S.
When do generalized entropies apply? How phase space volume determines entropy
 {\it Europhys Lett} {\bf 2011} {\em 96}, 50003.

\bibitem{ebook} 
 Thurner S.; Hanel R.
What do generalized entropies look like? An axiomatic approach for complex, non-ergodic systems.
In {\em Recent advances in Generalized Information Measures and Statistics}, 
Kowalski A.M.; Rossignoli R.;  Curado E.M.F., Eds.;  
Bentham Science eBook, in production 2013.

\bibitem{shannon}
Shannon C. E.  
A Mathematical Theory of Communication.
{\em The Bell System Technical Journal} {\bf 1948} {\em 27}, 379 and 623.

\bibitem{kinchin_1}
Khinchin A.I. 
{\em Mathematical foundations of information theory}.
Dover Publ., New York 1957. 

\bibitem{tsallis88}
Tsallis C.
{Possible generalization of Boltzmann-Gibbs statistics}. 
{\em J Stat Phys}  {\bf 1988} {\em  52}, 479-487.

\bibitem{celia}
Anteneodo C.;  Plastino A.R. 
Maximum entropy approach to stretched exponential probability distributions. 
{\em J Phys A: Math Gen} {\bf 1999} {\em 32}, 1089-1097. 

\bibitem{kaniadakis} 
Kaniadakis G.
{Statistical mechanics in the context of special relativity}. 
{\em Phys Rev E} {\bf 2002} {\em 66}, 056125.

\bibitem{curado}
Curado E.M.F.; Nobre F.D.
On the stability of analytic entropic forms. 
{\em Physica A} {\bf 2004} {\em 335}, 94-106.

\bibitem{expo_ent} 
Tsekouras G.A.; Tsallis C. 
Generalized entropy arising from a distribution of $q$ indices.   
{\em Phys Rev E} {\bf 2005} {\em 71}, 046144.

\bibitem{ggent}
Hanel R.; Thurner S. 
Generalized Boltzmann factors and the maximum entropy principle: entropies for complex systems.
{\em Physica A} {\bf 2007} {\em 380}, 109-114.

\bibitem{Hanel2011_b} 
Hanel R.; Thurner S.; Gell-Mann M.
Generalized entropies and the transformation group of superstatistics.
 {\em PNAS}  {\bf 2011}  {\em 108}, 6390-6394. 
 
 \bibitem{Hanel2012_b} 
Hanel R.; Thurner S.; Gell-Mann M.
Generalized entropies and logarithms and their duality relations.
 {\em PNAS} {\bf 2012}  {\em 109}, 19151-19154. 

\bibitem{TsallisBook_2009} 
Tsallis C. 
{\em Introduction to Nonextensive Statistical Mechanics}. 
Springer, New York 2009.

\bibitem{shafee}
Shafee F.   
Lambert function and a new non-extensive form of entropy.
{\em IMA J Appl Math} {\bf 2007} {\em 72}, 785-800. 

\bibitem{HT_hilhorst}
Hanel R., Thurner S.,  
Generalized-generalized entropies and limit distributions. 
{\em Braz J Phys} {\bf 2009} {\em 39}, 413-416. 

\bibitem{lesche}
Lesche B. 
Instabilities of R\'enyi entropies. 
{\em J Stat Phys} {\bf 1982} {\em 27}, 419-422.

\bibitem{Abe_2002}
Abe S.  
Stability of Tsallis entropy and instabilities of R\'enyi and normalized Tsallis entropies. 
{\em Phys Rev E}  {\bf 2002} {\em 66}, 046134.

\bibitem{Arimitsu} 
Jizba P.;  Arimitsu T.  
Observability of R\'enyiÕs entropy. 
{\em Phys Rev E} {\em 2004} {\em 69}, 026128. 

\bibitem{Kaniadakis_2004} 
Kaniadakis G.;  Scarfone A.M. 
Lesche stability of $\kappa$-entropy. 
{\em Physica A} {\bf 2004} {\em 340}, 102-109. 

\bibitem{HT_robustness_1}
Hanel R.;  Thurner S.; Tsallis C. 
On the robustness of q-expectation values and R\'enyi entropy.  
{\em Europhys Lett} {\bf 2009} {\em 85}, 20005.

\bibitem{TMGMS}
Tsallis C.; Gell-Mann M.; Sato Y. 
Asymptotically scale-invariant occupancy of phase space makes the entropy $S_q$ extensive
{\em PNAS} {\bf 2005}  {\em 102}, 15377-15382. 

\end{thebibliography}
\makeatletter
\renewcommand\@biblabel[1]{#1. }
\makeatother

\end{document}